\documentclass[twoside,11pt]{article}

\usepackage{obs_study_style}
\usepackage{siunitx}
\usepackage{amsmath}
\usepackage{graphicx}
\usepackage{url}
\usepackage{dcolumn,booktabs}
\usepackage{multirow}
\usepackage{bm}
\usepackage{enumitem}
\usepackage{comment}
\usepackage{xcolor}
\newcolumntype{d}[1]{D{.}{.}{#1}}
\newcommand{\bfX}{X}

\newcommand{\bfx}{x}

\heading{00}{00}{00}{00}{00}{Fan Li and Fan Li}

\ShortHeadings{Propensity scores methods for racial disparities}{Li and Li}
\firstpageno{1}

\begin{document}

\title{Using propensity scores for racial disparities analysis}
% Using propensity scores for racial disparities

\author{\name Fan Li \email fl35@duke.edu \\
       \addr Department of Statistical Science\\
       Duke University\\
       Durham, North Carolina, 27708, USA
  \AND
\name Fan Li \email fan.f.li@yale.edu \\
       \addr Department of Biostatistics \\
       Yale University\\
New Haven, Connecticut, 06510, USA
       }

\maketitle

\begin{abstract} 
\noindent Propensity score plays a central role in causal inference, but its use is not limited to causal comparisons. As a covariate balancing tool, propensity score can be used for controlled descriptive comparisons between groups whose memberships are not manipulable. A prominent example is racial disparities in health care. However, conceptual confusion and hesitation persists for using propensity score in racial disparities studies. In this commentary, we argue that propensity score, possibly combined with other methods, is an effective tool for racial disparities analysis. We describe relevant estimands, target population, and assumptions. In particular, we clarify that a controlled descriptive comparisons require weaker assumptions than a causal comparison. We discuss three common propensity score weighting strategies: overlap weighting, inverse probability weighting and average treatment effect for treated weighting. We further describe how to combine weighting with the rank-and-replace adjustment method to produce racial disparity estimates concordant to the Institute of Medicine's definition. The method is illustrated by a re-analysis of the Medical Expenditure Panel Survey data.
\end{abstract}

\begin{keywords}
covariate balance; propensity score weighting; racial disparities; target population; controlled descriptive comparison 
\end{keywords}

\section{Introduction}
Since first introduced in the landmark \citet{rosenbaum1983central} paper, propensity score has become a central concept in the potential outcome framework for causal inference. Following the dictum ``\emph{no causation without manipulation}'' \citep{holland1986statistics}, a ``cause'' refers to a \emph{treatment} or \emph{intervention} that is at least hypothetically manipulable. Hence for many researchers, the concept of propensity score is unequivocally tied with causal inference for manipulable interventions. However, a key property of the propensity score, the balancing property, does not involve potential outcomes nor imply causal interpretation. In fact, propensity scores have already been used in several non-causal contexts, e.g. generalizability \citep{stuart2011use}, constructing external controls in clinical trials from real world data \citep{lim2018minimizing}, covariate adjustment in randomized trials \citep{zeng2021propensity}. But a prominent exception is racial disparity studies, where the comparison is between different racial groups. Because race is not manipulable, racial disparity investigations are inherently not causal \citep{holland2003causation, Zaslavsky2005}, and thus there is often hesitation among health service researchers to use propensity scores for studying racial disparities. 

In this commentary, we aim to elucidate the basis and conditions for using propensity score for non-causal comparisons. The central message is that propensity score, as is defined in the broad sense, is a one-dimensional summary of the collection of pre-treatment covariates, and can be used as a numerical tool for balancing covariates between different groups, regardless whether the group assignment is (hypothetically) manipulable. Given the persistent misconception about propensity scores in racial disparities analysis, below we focus on the context of racial disparities in health care utilization. We highlight the importance of specifying target population \emph{a priori} and separating operational properties and contextual interpretation in propensity score analyses. We stress that our perspective is purely descriptive, and thus shall not be confused with the literature that addresses causal interpretation of racial comparisons, as in \cite{VanderWeele2014a,VanderWeele2014b} and the references therein.

\section{Controlled descriptive comparisons: estimands, target population and assumptions}

Let $Z$ denote the group membership variable, which is assumed to be binary for simplicity (e.g. White versus non-White), $\bfX$ denote a set of measured covariates, and $Y$ the observed health care outcome of interest. Because the goal is descriptive rather than causal comparisons, ``assignment'' here refers to a nonmanipulable state defining membership in one of two groups or populations, and the objective is a controlled comparison of the observed outcomes between the groups after adjusting for the differences in a set of pre-treatment covariates. Because confounding is a term closely tied to causal inference, we choose the term \emph{controlled} to emphasize the descriptive nature of the comparison. The propensity score $e(x)=\Pr(Z=1\mid \bfX=x)$ is the conditional probability of being in group $Z=1$ given covariates $\bfX$. 

Following \citet{li2013propensity}, we define the conditional average controlled difference (ACD) given covariate value $\bfx$ as,
\begin{equation}
\tau(\bfx)\equiv E(Y\mid Z=1,\bfX=\bfx)-E(Y\mid Z=0,\bfX=\bfx). \label{eq:ACD}
\end{equation}
%The quantity $\tau(\bfx)$ measures the net difference in the observed outcome between the two groups after controlling for differences in the observed covariates level $X=\bfx$. %In theory, there are infinity many covariate level over the continuous space of  $\bfX$.
Then we can define the average controlled difference on a scientifically meaningful target population by averaging the conditional ACD over that population, which can be represented as a weighted average controlled difference (WACD). Specifically, assume the observed sample is drawn from a population with probability density of covariates $f(X)$. Let $g(X)$ denote the covariate density of a pre-specified target population, which may be different from $f(X)$. We call the ratio $h(X)=g(X)/f(X)$ the \emph{tilting function}, which re-weights the distribution of the observed sample to represent the target population. Then we can represent the average controlled difference on the target population $g$ by a WACD estimand:
\begin{equation}\label{eq:tauh}
\tau_h=E_g[\tau(X)]=\frac{E[h(X)\tau(X)]}{E[h(X)]}.    
\end{equation}
%Assuming the marginal density of the covariates, $f(\bfX)$, exists, with a respect to a base measure $\mu$ (a product of counting measure with respect to categorical variables and Lebesgue measure for continuous variables).
%Given a pre-specified tilting function $h(\bfX)$, we can represent the target population density by $f(\bfX)h(\bfX)$, which leads to the WACD estimand
% \begin{equation}\label{eq:tauh}
% \tau_h=\frac{\int_{\bfx\in\mathcal{X}}\tau(\bfx)f(\bfx)h(\bfx)\mu(d\bfx)}{\int_{\bfx\in\mathcal{X}}f(\bfx)h(\bfx)\mu(d\bfx)}.
% \end{equation}
In plain language, $\tau_h$ refers to the average net difference in the health care outcome $Y$ between two groups with their covariate distributions adjusted to be the same as in the target population $g$. Different tilting functions lead to different target population, as illustrated in the following three examples. For notational purposes, we define $\mathbb{S}_0$ and $\mathbb{S}_1$ as the units in the $Z=0$ group and the $Z=1$ group, respectively. %(or proportional up to a constant) 
\begin{itemize}

\item When $h(\bfX)\propto 1$, $\tau_h$ represents the average difference in the outcome $Y$ once we force the distribution of $\bfX$ in each of $\mathbb{S}_0$, $\mathbb{S}_1$ to be identical to that in the union $\mathbb{S}_0\cup \mathbb{S}_1$, which is the target population. In the racial disparity context, this target population would be the overall population combining the two racial groups under comparison with all covariates being balanced.   

\item When $h(\bfX)\propto e(\bfX)$, $\tau_h$ represents the average difference in the outcome $Y$ once we force the distribution of $\bfX$ in group $\mathbb{S}_0$ to be identical to that in group $\mathbb{S}_1$. The same argument applies to $h(\bfX)\propto \{1-e(\bfX)\}$ if we flip the definition of $Z=1$ and $Z=0$. So the target population is group $\mathbb{S}_1$ or $\mathbb{S}_0$. In the racial disparity context, it would be a population with the same covariate distribution as one pre-specified racial group. 

\item When $h(\bfX)\propto e(\bfX)\{1-e(\bfX)\}$, $\tau_h$ represents the average difference in the outcome $Y$ once we force the distribution of $\bfX$ in each of $\mathbb{S}_0$, $\mathbb{S}_1$ to be identical to that in the subpopulation that has the largest tendency to belong to both $\mathbb{S}_0$ and $\mathbb{S}_1$, a concept similar to equipoise in clinical evaluations. We call this target population the overlap population \citep{li2018balancing,LiThomasLi2018,cheng2022addressing}, which is akin to an intersection---rather than a union---between $\mathbb{S}_0$ and $\mathbb{S}_1$. In the racial disparity context, this target population would be the subpopulation with the most similar covariate distribution between the two racial groups.% Effectively the target population is this overlap population, within which we are making descriptive comparisons.
\end{itemize}

We stress a key but under-appreciated distinction between causal and controlled descriptive comparisons in terms of necessary assumptions, namely, the latter requires weaker assumptions than the former. Causal comparisons typically require (A1) the Stable Unit Treatment Value Assumption (SUTVA), which states there is no interference between units and no different versions of the treatment; (A2) the unconfoundedness assumption, which states there is no unmeasured confounder; and (A3) the overlap or positivity assumption, which states that each unit has non-zero probability of being in either group. SUTVA underpins the existence of two potential outcomes for each unit; uconfoundedness connects potential outcomes to the observed outcomes and enables interpreting the difference in the observed outcomes as a causal effect. However, a controlled descriptive comparison merely needs to adjust for the difference in the covariates between two groups rather than offers a causal interpretation of the difference in the observed outcome. Therefore, it does not involve the concept of potential outcomes and consequently does not require SUTVA or unconfoundedness. On the other hand, the overlap assumption is necessary for both causal and controlled descriptive comparisons. This is first for an operational reason because propensities close to 0 and 1---equivalently lack of overlap---lead to large variance in estimation. Moreover, overlap is needed in causal comparisons for an additional conceptual reason to define causal effects: if a unit has zero probability to be assigned to one group, then we cannot conceive its potential outcome corresponding to that group assignment. 

\section{Propensity score methods for estimation}
We now outline two estimation methods of the WACD: matching and weighting. In matching, one chooses an algorithm finding pairs of units in two groups with similar covariates according to some distance metric (e.g. the propensity score), and then calculates the difference in the average observed outcome between the groups in the matched sample \citep{rubin2006matched}. The overlap assumption is in effect achieved by dropping the unmatched units. Matching is a bottom-up method in the sense that it starts from local balance in the observed sample rather than global balance in a pre-defined target population. The target population is defined only implicitly through the matching algorithm. For example, when the matching algorithm is designed to find matches for (nearly) all units in the whole sample, e.g. the full matching method \citep{rosenbaum1991characterization, hansen2004full}, the target population is the overall population. When the algorithm is designed to find matches for (nearly) all units from one specific group (e.g. the treated group in the causal context or a racial group in the disparities context), the target population is the population in that group, corresponding to $\tau_h$ with $h(X)=e(X)$. When the matching algorithm is designed to find matches for a subset of marginal units, that is, units who might or might not belong to a specific group, e.g. the optimal matching method \citep{rosenbaum2012optimal}, the target population is the overlap population, corresponding to $\tau_h$ with $h(X)=e(X)\{1-e(X)\}$. 

Weighting methods assign a weight to each unit and then calculate the weighted difference in outcomes between the comparison groups. Weighting is a top-down method in the sense that it starts from the global balance in a target population rather than local balance in pairs in the sample. The target population is explicitly pre-specified and determines the corresponding weighting scheme (as elaborated below). Unlike matching, there is no automatic procedural guarantee of the overlap assumption in weighting, which may cause inflated variances in some weighting schemes such as the inverse probability weighting. Below we will focus on the weighting method because it more directly connects to the definition of the WACD estimands.  % and in what follows, the overlap assumption is needed only for the purpose of defining the balancing weights.

% both causal and descriptive comparisons, but for different purposes. 
%For example, without overlap, the inverse probability weighting estimator will have extreme weights and inflated variance, whereas matching estimators will drop a large number of units.  Notice that the overlap assumption is not needed so that we can define both potential outcomes, which are only needed for causal comparisons. 
%This is a subtle but important difference because it has not been widely appreciated that descriptive comparisons require weaker assumptions. 

For any pre-specified target population $g(X)$ and equivalently the tilting function $h(X)$, we define the corresponding balancing weights \citep{li2018balancing}: 
\[ \left\{ \begin{array}{ll}
w_1(\bfX)\propto{h(\bfX)}/{e(\bfX)},  & \mbox{for } Z=1\\
w_0(\bfX)\propto{h(\bfX)}/{(1-e(\bfX))}, &\mbox{for } Z=0,
\end{array} \right.\]
It is straightforward to show that the WACD estimand \eqref{eq:tauh} can be represented as the weighted difference in the mean outcome between the two groups
\begin{align*}
\tau_h=\frac{\mathbb{E}[YZw_1(\bfX)]}{\mathbb{E}[Zw_1(\bfX)]}-\frac{\mathbb{E}[Y(1-Z)w_0(\bfX)]}{\mathbb{E}[(1-Z)w_0(\bfX)]},
\end{align*}
without invoking SUTVA or the unconfoundedness assumption. To estimate WACD, one can then  consider the following Haj\'ek estimator
\begin{align}
\widehat{\tau}_h=\frac{\sum_{i=1}^N Y_iZ_iw_1(\bfX_i)}{\sum_{i=1}^N Z_iw_1(\bfX_i)}-\frac{\sum_{i=1}^N Y_i(1-Z_i)w_0(\bfX_i)}{\sum_{i=1}^N (1-Z_i)w_0(\bfX_i)}, \label{eq:hajek}
\end{align}
where the weights $\{w_1(\bfX_i),w_0(\bfX_i)\}$ are based on estimated propensity scores $\widehat{e}(\bfX_i)$, e.g. via a logistic regression. 

Here we list the balancing weights corresponding to previously discussed three target populations. When $h(X_i)\propto 1$, $\{w_1(\bfX_i)=1/e(X_i),w_0(\bfX_i)=1/\{1-e(X_i)\}\}$ is the inverse probability weight (IPW). Conceptually, IPW balances the covariates toward the overall population represented by the sample. Operationally, IPW over-weight units who have large probability to being in the opposite group (i.e. with propensity scores close to 0 for units in $\mathbb{S}_1$ or close to 1 for units in $\mathbb{S}_0$). In other words, these units' characteristics are the least similar to the opposite group. Such a feature is not desirable in the context of racial disparities because arguably we want to over-weight the units who are the most similar between groups (i.e., with propensity scores close to 0.5). This is exactly what overlap weight (OW), $\{w_1(\bfX_i)= \{1-e(\bfX_i)\}, w_0(\bfX_i)= e(\bfX_i)\}$  (corresponding to $h(X_i)\propto e(X_i)\{1-e(X_i)\}$), is designed to achieve. Conceptually,  OW emphasizes a naturally comparable subpopulation with similar health status, namely patients whose race category, conditional on their health conditions and clinical need, are indistinguishable. Operationally, \cite{li2018balancing} showed that the overlap weights lead to the smallest asymptotic variance of $\hat{\tau}_h$ among balancing weights. Moreover, when the propensity score is estimated by a logistic regression, the resulting overlap weights lead to exact mean balance of any covariate $X$ included in the regression. Therefore, the overlap weights remove all imbalances in measured covariates $\bfX$ among the overlap population, rendering it particularly suitable for controlled descriptive comparisons.  When $h(X_i)\propto e(X_i)$, $\{w_1(\bfX_i)=1,w_0(\bfX_i)=e(X_i)/\{1-e(X_i)\}\}$ is the so-called ATT (average treatment effect for the treated) weight. The operating characteristics of ATT weights are similar to those of IPW, but limit to one group.

\section{Application to racial disparities analysis in health care utilization}\label{sec:PSW}
\subsection{Propensity score weighting analysis} \label{sec:PSW_MEPS}
The \emph{Unequal Treatment} report from the Institute of Medicine (IOM) defines health care disparity as the difference in treatment provided to social groups that is not justified by health status or treatment preference of the patient \citep{IOM2003}. To be concordant with the IOM definition, analysts need to adjust for the health status variables across different racial groups in disparity studies \citep{Cook2012}. 

\begin{figure}[htbp]
\centering
\includegraphics[scale=0.55]{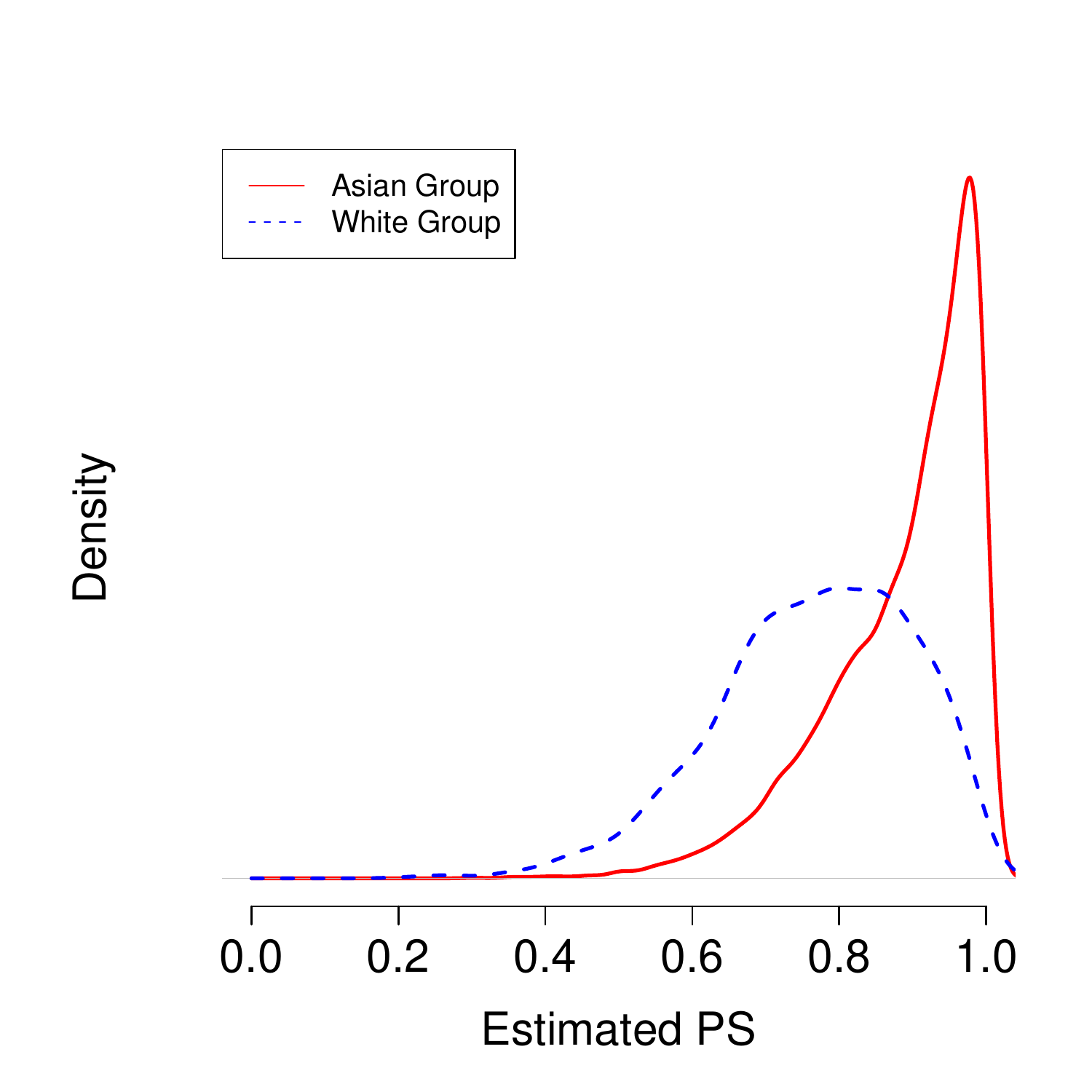}
\caption{Distribution of the estimated propensity scores for being in the White group.}
\label{fig:Fig1}
\end{figure}

\begin{figure}[htbp]
\centering
\includegraphics[scale=0.5]{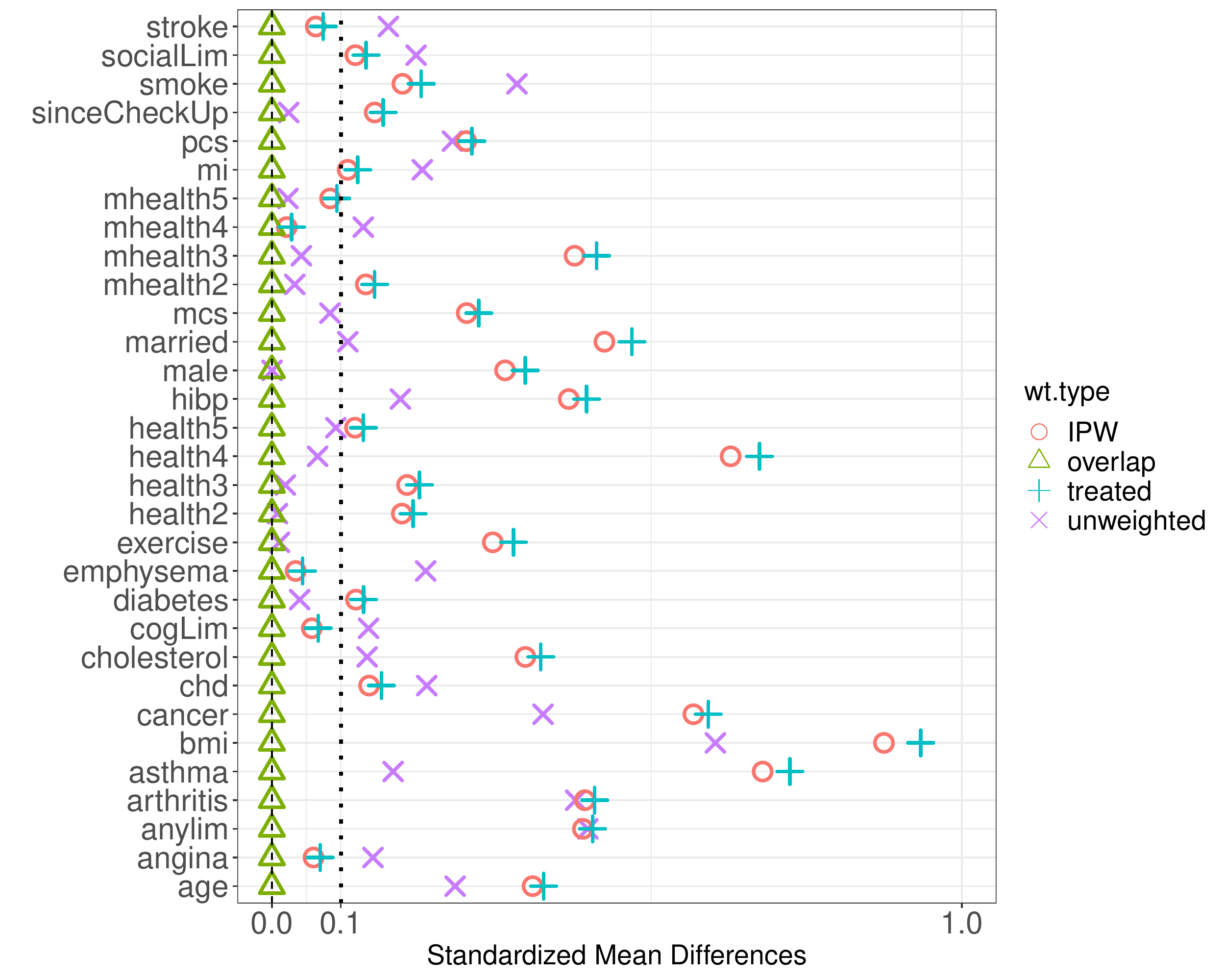}
\caption{Love plot of the absolute standardized difference for each covariate in the original and weighted data. The dotted vertical line indicates the threshold for balance at $0.1$. IPW: inverse probability weighting; overlap: overlap weighting; treated: ATT weighting.}
\label{fig:Fig2}
\end{figure}

For illustration, we apply propensity score weighting to the 2009 Medical Expenditure Panel Survey (MEPS) to study the White-Asian disparities in medical expenditure. The analysis is implemented using the R package \texttt{PSweight} \citep{zhou2022Rjournal}. Details on other White-minority such as White-Black and White-Hispanic comparisons can be found in \cite{Cook2010} and \cite{li2019propensity}. The sample contains 9830 non-Hispanic White and 1446 Asian adults aged at least 18 years. Here the propensity score is the probability of being in the White group, estimated from a logistic propensity score model controlling for the following health status variables (denoted by $\bfX_H$): body mass index, SF-12 physical and mental component summary, self-reported health status, measurements of health conditions including diabetes, blood pressure, asthma, MI, stroke, age, gender, and marital status. Figure \ref{fig:Fig1} presents the distribution of the estimated propensity scores (for being in the White group) by each observed group. There is a substantial proportion of units, particularly Asians, having propensities close to 1, suggesting a lack of overlap. We applied IPW, ATT weighting and OW to balance the covariates. The resulting weighted covariate balance, measured by the Absolute Standardized Difference (ASD), is displayed in Figure \ref{fig:Fig2} for each covariate. A rule of thumb for covariate balance in the literature is ASD smaller than $0.1$ \citep{austin2015moving}. Clearly, both IPW and ATT lead to insufficient balance in a number of covariates with the largest ASD close to 1, whereas OW leads to identically zero ASD in all covariates, effectively removing any difference in the outcomes that is attributable to the health status variables in the overlap population.

The estimated disparities in total health expenditure based on different weighting methods are presented in Table \ref{tb:result}. The standard errors are estimated using the closed-form sandwich variance estimator in \citet{LiThomasLi2018}, which accounts for the uncertainty in estimating the propensity scores. Table \ref{tb:result} shows that Whites spent \$2167 (95\% CI (117, 4217)) , \$2310 (95\% CI (244, 4376)), \$1227 (95\% CI (796, 1658)) more on health care than Asians, using IPW, ATT and OW, respectively. This shows that disparity estimates may be sensitive to the choice of target population used for balancing covariates. The White-Asian disparity estimates from IPW and ATT are likely subject to bias because IPW and ATT fail to adequately balance the health status variables, as shown in Figure \ref{fig:Fig2}. Besides, the lack of overlap leads to much inflated standard errors in IPW and ATT, both of which are five times of that of OW. %A more statistically efficient disparity estimate was identified under OW, as expected from the variance minimizing property proved in \citet{li2018balancing}.

\begin{table}[htbp]
\centering
\caption{White-Asian disparities in total health care expenditure (in dollars). The point estimates are obtained as average controlled differences (before and after propensity score weighting). IPW: inverse probability weighting; ATT: ATT weighting; OW: overlap weighting; IOM-c: IOM definition concordant.}\label{tb:result}\vspace{0.1in}
\begin{tabular}{lccccccc}
\toprule
 & Unweighted & IPW & IPW (IOM-c) & ATT & ATT (IOM-c) & OW & OW (IOM-c)  \\ \midrule
Estimate & $2764$  & $2167$  & $-933$ & $2310$ & $-1011$ & $1227$  & $1063$ \\
SE &$225$ & $1046$ & $1963$  & $1054$ & $2046$ & $220$ & $238$  \\
\bottomrule
\end{tabular}
\end{table}

\subsection{Combine propensity score weighting with rank-and-replace adjustment}
The IOM definition of disparity includes racial differences in utilization mediated through factors other than health status and preference, such as many social factors \citep{McGuire2006}. This renders the analyses that only adjust for health status characteristics as in Section \ref{sec:PSW_MEPS} inadequate. A number of methods have been developed to adjust for socioeconomic status (SES) information in racial disparities studies in health services \citep[e.g.][]{McGuire2006, Cook2009}. In particular,  \citet{McGuire2006} advocate to distinguish between health status variables ($\bfX_H$) and SES variables ($\bfX_S$) in analysis. 
If the health status variables $X_H$ are correlated with the SES variables $X_S$,  propensity score weighting adjusting for $X_H$ (as that in Section \ref{sec:PSW_MEPS}) may inadvertently alter the distributions of $X_S$ and only provide an approximation to the IOM-defined disparity \citep{Balsa2007}. \cite{McGuire2006} developed the rank-and-replace adjustment method to undo the undesired weighting of $\bfX_S$ due to its correlation with $\bfX_H$. 
%
%we cannot specify any variable as preference measurements, but acknowledge that the lack of this information represents a limitation in implementing the IOM definition in racial disparities.
Below we combine the rank-and-replace adjustment with our propensity score analysis of MEPS as a further illustration. Here the SES variables include poverty status, education, health insurance and geographical region. %We perform the rank-and-replace adjustment based on a model-based SES index to equalize the weighted SES distributions and the unweighted marginal distributions \comment{of what?}. 
We impose a log-linear model to model the total health care expenditure as a function of $\bfX_H$, $\bfX_S$ and the racial group indicator 
\begin{align}\label{eq:outcome}
\log(\mathbb{E}[Y_i|\bfX_{H,i},\bfX_{S,i},Z_i])=\gamma_0+\gamma_{1}Z_{i}+\bfX_{H,i}^T\gamma_H+\bfX_{S,i}^T\gamma_S.
\end{align}
We take the fitted value of $\bfX_{S,i}^T\gamma_S$ as the individual SES predictive index, %For health expenditure data, we follow \citet{Buntin2004} to allow for heteroscedastic variances, and apply the Park test to determine the variance power relative to the mean \citep{Park1966,Manning2001}. \comment{Do we need the above detail? I feel it is not important and too similar to AOAS} 
and first obtain the propensity score weighted (according to each specific weighting scheme) rank of $\bfX_{S,i}^T\gamma_S$ within each race. To restore the original group-specific SES distributions, we then replace $\bfX_{S,i}^T\gamma_S$ for unit $i$ with $\bfX_{S,j}^T\gamma_S$ for unit $j$ such that the weighted rank of unit $i$ equals the unweighted rank of unit $j$ within each racial group. With this adjustment, the weighted distribution of the SES index in each group is approximately the same as the original distribution of the SES index in that group. We then predict the expenditure outcome for each individual based on model \eqref{eq:outcome}, after the rank-and-replace adjustment, and estimate the WACD using this predicted expenditure outcome. The resulting disparities estimates become more IOM-concordant in the sense that they recapture the racial differences in SES even after propensity score weighting of the health status variables. The final disparity estimates are obtained by re-weighting (e.g. via OW or IPW) the predicted total health care expenditure in model \eqref{eq:outcome} after the rank-and-replace adjustment. % to restore the original group-specific SES distributions. 
% need to acknowledge that this approach is model dependent, a bit similar to DR, but also fundamentally different, the only similarity is that we need an outcome model

Table \ref{tb:result} provides the disparity estimates by combining propensity score weighting with rank-and-replace adjustment (labeled as IOM-c); the standard error estimates are obtained based on $1000$ bootstrap replicates. Interestingly, these estimates are more sensitive to the choice of the target population than the original weighting analysis. This is particularly noticeable with IPW. Specifically, IPW suggests that Whites spent on average \$933 less than Asians when only adjusting for the difference in health status (but not the SES). As implied from the lack of balance in Figure \ref{fig:Fig1} and by numerous simulation studies \citep{LiThomasLi2018,li2019propensity}, the IPW disparities estimates are likely subject to bias. In comparison, the disparities estimates under OW appear more stable. In particular, when we only address differences in health status variables, OW suggests that Whites spent on average \$1063 more on health care than Asians, after using rank-and-replace adjustment to restore differences due to SES variables.

Finally, we note that one limitation in implementing the IOM concordant disparities analysis, even after the rank-and-replace adjustment, is that there is no consensus on how to measure patient preferences. More substantive guidance on this would benefit racial disparities studies in health care.    

\section*{Acknowledgement}
The authors are grateful to Elizabeth Stuart for initiating this topic in a conversation, to Alan Zaslavsky and Peng Ding for insightful discussions, to Chao Cheng for computational assistance in the MEPS analysis in Section \ref{sec:PSW}.
\vskip 0.2in
\bibliography{OS_comment}

\end{document}